\documentclass[11pt]{article}
\usepackage{jheppub}
\usepackage{amsmath}
\usepackage{amsfonts}
\usepackage{amssymb}
\usepackage{mathrsfs}
\usepackage{graphicx}
\usepackage{multirow}
\usepackage{color}
\usepackage{hyperref}
\usepackage{caption}
\usepackage{subcaption}

\usepackage{slashed}
\newcommand{\lang}{\mathcal{L}}

\def\beq{\begin{equation}\begin{aligned}}
\def\eeq{\end{aligned}\end{equation}}

\def\inv{^{\raise.15ex\hbox{${\scriptscriptstyle -}$}\kern-.05em 1}}
\def\lbar{{\lower.35ex\hbox{$\mathchar'26$}\mkern-10mu\lambda}} 

\let\<=\langle
\let\>=\rangle

\let\+=\uparrow

\definecolor{orange}{rgb}{1,0.5,0}

\begin{document}
\title{Heavy superpartners with less tuning from hidden sector renormalisation}


\author{Edward Hardy}
\emailAdd{e.hardy12@physics.ox.ac.uk}
\affiliation{Rudolf Peierls Centre for Theoretical Physics, University of Oxford,\\
1 Keble Road, Oxford, OX1 3NP, United Kingdom}


\abstract{
In supersymmetric extensions of the Standard Model, superpartner masses consistent with collider bounds typically introduce significant tuning of the electroweak scale. We show that hidden sector renormalisation can greatly reduce such a tuning if the supersymmetry breaking, or mediating, sector runs through a region of strong coupling not far from the weak scale. In the simplest models, only the tuning due to the gaugino masses is improved, and a weak scale gluino mass in the region of 5 TeV may be obtained with an associated tuning of only one part in ten. In models with more complex couplings between the visible and hidden sectors, the tuning with respect to sfermions can also be reduced. We give an example of a model, with low scale gauge mediation and superpartner masses allowed by current LHC bounds, that has an overall tuning of one part in twenty.
}

\maketitle


\section{Introduction}
Supersymmetry (SUSY) provides a compelling solution to the electroweak (EW) heirachy problem of the Standard Model (SM). However, ever increasing limits on superpartners from the LHC are very challenging to accommodate without introducing significant fine tuning. Even extensions of the simplest models, such as the next to minimal supersymmetric model (NMSSM), natural SUSY, R-parity violation, and Dirac gauginos cannot completely rescue the situation and the theories obtained are typically fine tuned to at least one part in $100$.

In this paper, we show that the fine tuning with respect to the mass of the superpartners, at the assumed ultraviolet (UV) boundary of the renormalisation group (RG) flow, can be substantially reduced through the effects of hidden sector renormalisation, first studied in \cite{Nelson:2000sn,Dine:2004dv,Cohen:2006qc}, and later expanded on in \cite{Murayama:2007ge,Abe:2007ki,Cho:2008fr,Kawamura:2008bf,Campbell:2008tt,Perez:2008ng,Craig:2009rk,Arai:2010ds,Arai:2010qe}. This is an effect where the RG  flow of MSSM scalar soft masses may be significantly modified by the details of the hidden SUSY breaking sector. In particular, if the SUSY breaking sector runs through a region of strong coupling close to a conformal fixed point, and the operator coupling the Higgs superfields to the SUSY breaking spurion obtains a large anomalous dimension, it can efficiently suppress the Higgs soft mass. This washes out the dependence of the EW scale on the superpartner masses and as a result the fine tuning of phenomenologically viable theories can be significantly reduced.

More precisely, suppose the operator that generates the Higgs soft mass gains a large anomalous dimension between energy scales $\Lambda_1$ and $\Lambda_2$. Then, approximating the anomalous dimension, $\gamma$, as a constant in this region, after the strong coupling region the Higgs soft mass is given by
\begin{equation}
m_{Hu}^2\left(\Lambda_{1}\right) \approx m_{Hu}^2\left(\Lambda_{2}\right) \left(\frac{\Lambda_1}{\Lambda_2}\right)^{\gamma} .\label{eq:1}
\end{equation}
Any feed in to the Higgs mass from superparticles above, or during, the strong coupling region is strongly suppressed if the strong coupling regime lasts for a relatively long time. Provided the strong coupling ends not far from the weak scale, there is little time for a dependence on the superparticle masses to reemerge, and the fine tuning is dramatically reduced.\footnote{The possibility that the Higgs mass operator may obtain a large anomalous dimension reducing fine tuning in SUSY theories has previously been considered \cite{Terao:2001jw,Kobayashi:2004pu,Terao:2007pm,Cohen:2012rm}. The main difference in the models we study, is that the strong coupling and large anomalous dimensions arise directly in the SUSY breaking or mediation sector, rather than some additional sector coupled to the Higgs.}

Rather than attempting to study particular examples of strong dynamics explicitly, which is both notoriously difficult and may not capture generic features of such sectors, we simply parameterise the effect of the strong coupling region by assuming certain operators get large anomalous dimensions in this region. Of course, the lack of an example of a theory combining all the required elements is a significant deficiency of our present work. However given that only a handful of models of dynamical supersymmetry breaking are known, and even fewer are actually calculable, this is perhaps acceptable. Later we argue that it is possible that sectors with the appropriate dynamics can exist and  describe models which exhibit some of the required features.

In the simplest implementations, all chiral multiplets are assumed to couple universally to the hidden sector, and the soft mass operators obtain equal anomalous dimension during the strong coupling period. Due to the universal couplings,  the sfermion masses, as well as the Higgs mass, are suppressed during the RG flow. Obtaining weak scale sfermion masses in the region of several TeV requires them to be heavier than normal at the UV boundary of the RG flow, and so the tuning with respect to these states is not reduced. If the hidden sector is approximately supersymmetric during the strong coupling region, the gaugino mass operator is protected from gaining a large anomalous dimension by non-renormalisation theorems and holomorphy. Even if the sector is non-supersymmetric, it is quite plausible that models exist where this operator does not obtain a large anomalous dimension since it is distinguished from the operators that generate scalar soft masses, for example due to its R-symmetry breaking nature. Therefore, the tuning with respect to gaugino masses is substantially reduced.  Due to large production cross sections and dramatic decay signals, there are stringent collider limits on the gluino mass. A large gluino mass feeds strongly into the Higgs mass through the stops during RG flow, so this is often the dominant tuning in a theory \cite{Hardy:2013ywa,Arvanitaki:2013yja}, and even these most basic models of hidden sector renormalisation can be a substantial improvement over traditional theories.

More complex models with extra interactions between the visible and hidden sector can reduce the fine tuning with respect to the sfermion masses as well. For example, this can occur in a theory where the Higgs has extra couplings to the SUSY breaking sector. These may cause the Higgs soft mass operator to gain a large anomalous dimension, while the sfermion operators do not, reducing the dependence of the Higgs mass on the sfermions. Since the sfermion couplings remain universal, strong constraints from flavour observables that are often challenging to accommodate in SUSY models are satisfied. An even more exotic possibility is that the Higgs soft mass operator obtains an opposite sign anomalous dimension to the sfermion mass operators. This leads to an enhancement of the sfermions masses while the Higgs mass is still suppressed. Potentially, soft masses in the region of 10 TeV can be obtained without making the usual tuning of approximately 1 part in 100 any worse.

As well as the SUSY particles' soft masses, the fine tuning of the EW vacuum expectation value (VEV) depends on the $\mu$ and $B\mu$ parameters at tree level. To obtain low fine tuning, these must be relatively small. Since the LHC is fairly insensitive to charginos, this is not an very severe constraint. It does however open up the prospect that a future collider may discover light charginos, with other superpartners potentially much heavier. A further attractive possibility for future work would be to build a model where appropriately sized  $\mu$ and $B\mu$ terms are generated through hidden sector renormalisation (as has been previously studied) while simultaneously explaining why only the Higgs mass operator gains a large anomalous dimension. For the majority of our study we consider traditional Majorana gauginos which feed into the sfermions at all energy scales. Later we briefly comment on the interesting extension of the MSSM to (supersoft) Dirac gauginos \cite{Fox:2002bu}, which can reduce the gaugino fine tuning even further, although the tunings with respect to the $\mu$ parameter and initial Higgs soft mass are unchanged.

Turning to the structure of the paper, we begin in Section \ref{sec:2} by briefly reviewing hidden sector renormalisation. In Section \ref{sec:gaugino} we discuss the mechanism that reduces the fine tuning, and carry out a full numerical study of the fine tuning in models with hidden sector renormalisation. Section \ref{sec:model} contains a discussion of the types of theory that can may lead to the required dynamics and other model building possibilities, and we conclude in Section \ref{sec:con}.


\section{Hidden sector renormalisation}\label{sec:2}
The models we study are similar to those introduced in \cite{Dine:2004dv}, with the crucial difference that the region of strong coupling occurs close to the weak scale. Consider a SUSY breaking sector with a spurion $X$, which receives an F-term, $F_0$, at the UV boundary of the RG flow. The visible sector scalars and gauginos get mass through terms in the effective Lagrangian
\begin{equation} \label{eq:massop}
\lang \supset \int d^4\theta a_i \frac{X^{\dagger}X}{M^2_*} \Phi^{\dagger}_i\Phi_i +\int d^2\theta  w_n \frac{X}{M_*} W_{n\alpha}W_n^{\alpha} +\rm{h.c.} ,
\end{equation}
where $\Phi_i$ represents the visible sector chiral superfields, $W_n$ is the gauge field strength (corresponding to the gauge group $n$), and $M_*$ is some high energy scale in the theory. These terms may be generated by integrating out messenger fields in models of gauge mediation, or interactions with other heavy states.

We consider both models where the hidden sector is approximately supersymmetric during the strong coupling region and also models where SUSY is broken at this scale. If the hidden sector is supersymmetric, the holomorphic coupling $w_n$ is not renormalised perturbatively.\footnote{Non-perturbative renormalisation of holomorphic couplings is not forbidden by non-renormalisation theorems, and may be important in the strong coupling region. We assume that this does not lead to new operators that generate soft masses in the visible sector.} As a result, the physical gaugino mass only flows due to the wavefunction renormalisation of $X$, along with the standard flow of the gauge coupling. Denoting the wavefunction renormalisation of $X$ by $Z_X$, and normalising such that $Z_X=1$ at the UV boundary of the RG flow, the physical gaugino mass at a scale $\mu$ is
\begin{equation}
M_{n}\left(\mu\right) = g_n^2\left(\mu\right) w_n \frac{F\left(\mu\right)}{M_*} . 
\end{equation}
where we have defined $F\left(\mu\right) = \frac{F_0}{Z_X^{1/2}\left(\mu\right)}$. Here, $g\left(\mu\right)$ is the gauge coupling in a basis where all fields are canonically normalised. Its RG evolution is given by the NSVZ beta function,
\begin{equation}\label{eq:nsvz}
\beta\left(g\right) = -\frac{g^3}{16 \pi^2}\frac{3 T\left(Ad\right) - \sum_i T\left(R_i\right)\left(1-\gamma_i\right)}{1- \frac{g^2}{8\pi^2} T\left(Ad\right)} ,
\end{equation}
where $i$ labels the matter fields in the theory, which are in the representation $R_i$, and have anomalous dimension $\gamma_i$, and $T\left(R_i\right)$ is the Dynkin index of the representation $R_i$. This arises as a combination of a one loop exact renormalisation of the holomorphic gauge coupling, and a rescaling anomaly from canonically normalising the fields in the theory.

In contrast, the non-holomorphic operator that leads to scalar masses is renormalised. Crucially, this is separate, and in addition to, the wavefunction renormalisation of $X$. It is this renormalisation that means that the dynamics of the hidden sector do not simply result in a rescaling of all soft masses. Including the wavefunction renormalisation of $X$ through the rescaling of $F_0$, the scalar masses are
\begin{equation}
m^2_i = a_i\left(\mu\right) \frac{F\left(\mu\right)^2}{M_*^2} .
\end{equation}
Here $a_i\left(\mu\right)$ is the renormalised coupling, which evolves according to
\begin{equation}
\frac{da_i}{dt}= \tilde{\gamma}_i a_i - \frac{1}{16\pi^2} \sum_n 8C_n\left(R_i\right) g_n^6 w_n^2 + ... \, , \label{eq:rg}
\end{equation}
where $t= \log\mu$, and $C_n$ is the quadratic Casimir of the state $i$ with respect to the gauge group $n$. $\tilde{\gamma}_i$ is the extra contribution to the anomalous dimension of the operator from hidden sector effects beyond wavefunction normalisation of $X$, and three ellipses represents other visible sector one-loop effects, for example those proportional to the Yukawa couplings, and terms from higher loops.\footnote{For simplicity, we assume throughout that the SM states do not couple significantly to any hidden sector operators other than $X$ and $X^{\dagger}X$.  The extension to more general cases is straightforward \cite{Cohen:2006qc}.}

If the theory is non-supersymmetric during the strong coupling region, consistently packaging the fields of the theory into supermultiplets is no longer possible. Regardless, we will see in Section \ref{sec:model} that there are models where the operators that generate the visible sector masses may continue to be renormalised below the scale $\sqrt{F}$. Of course, the argument from holomorphy protecting the gaugino mass does not hold in this case. However, even if SUSY is broken, it seems plausible that there exist theories where the gaugino mass operator gains a far smaller anomalous dimension than the scalar mass operators. This is because the gaugino mass is an R-symmetry breaking operator and the interactions of a vector multiplet are necessarily different to those of a chiral multiplet.

While it is possible to study the effects of a particular model of the hidden sector, similarly to \cite{Arai:2010qe}, we take an alternative approach and \emph{parameterise} the impact of running through a strong coupling regime. This is done by turning on large anomalous dimensions for some operators in the energy region of strong coupling. Due to unitarity, physical operators have positive total anomalous dimension at one loop \cite{Higashijima:2003et}, however this requirement does not persist at higher orders and so does not apply during strong coupling.  For simplicity, we also assume that $Z_X\left(\mu\right)=1$ at all energy scales, so that the physical F-term of $X$ does not flow. This does not alter the phenomenology of our models, since it is the coefficients of the operators, $a_i$ and $w_n$, which feed into each others RG, not the masses. Making this assumption just means the RG flow of the soft masses is not rescaled relative to that of the couplings.


\section{Fine tuning in the presence of a strongly coupled hidden sector}\label{sec:gaugino}
The EW fine tuning of a theory, with respect to the parameter $i$, is defined as 
\begin{equation}
Z_i = \frac{\partial \log m_Z^2}{\partial \log i}. \label{eq:ft}
\end{equation}
We follow the convention that the tuning is measured with respect to masses squared (this introduces a factor of 2 in the tuning compared to that with respect to masses). In evaluating the fine tuning, throughout we assume that $\tan\beta$ is moderately large, so that the EW scale is well approximated as
\begin{equation}
m_Z^2 = -2 \left( m_{Hu}^2+\left|\mu\right|^2 \right) + \mathcal{O}\left(\frac{1}{\tan^2\beta} \right) . \label{eq:ew}
\end{equation}
Under this assumption, the down type Higgs mass is not significant for the fine tuning.\footnote{The extension to small $\tan\beta$ is straightforward, and does not alter our numerical results substantially.}

In order to capture important effects from running, the derivative in eq. \eqref{eq:ft} should be taken with respect to the value of the parameter at the UV cut-off of the theory. In the present work, we do not specify a UV completion of our models, therefore it is not possible to measure the fine tuning with respect to parameters at this scale, since to do so requires a complete knowledge of all higher dimensional operators. Instead, we simply assume some UV boundary condition to the RG flow, and vary the parameters, which we take to be independent, at this location. While this is only an approximation to the true tuning of a complete theory, it does highlight correlations generated by running that that may well remain important when embedded in a full theory.

An alternative measure of fine tuning is available, the tuning  with respect to the values of the parameters at the EW scale. As discussed in \cite{Baer:2013bba,Baer:2013gva}, this gives a lower bound on the fine tuning of a theory, avoiding assumptions about the UV completion of the low energy effective field theory. For example, correlations between soft parameters at the UV cutoff of the theory could mean that a theory's true fine tuning is much lower than a naive estimate of the tuning based on high scale parameters would suggest. Similarly, the models we study in this paper are examples of theories where assuming the RG flow is just that of the MSSM would lead to an overestimate of the high scale tuning. Effectively, hidden sector renormalisation can lower the tuning with respect to the high scale parameters towards the lower bound set by the tuning with respect to the weak scale parameters.

To obtain phenomenologically acceptable models with reduced fine tuning, the strong coupling region has to end not far from the weak scale (typically at a few TeV) so that a perturbation to the Higgs soft mass is not regenerated before the superparticles are integrated out of the effective theory. Additionally, the strong coupling must extend over at least roughly one order of magnitude in energy scale, so that the Higgs soft mass is sufficiently suppressed. This can occur if the RG flow passes very close to an interacting conformal fixed point. Depending on the details of the models involved, the theory may be either supersymmetric and non-supersymmetric during the strong coupling regime. If the SUSY breaking sector itself is responsible for the hidden sector renormalisation, the scale of mediation must be low, however if it is the messenger sector that runs to strong coupling near the weak scale the scale of mediation can be high. In Sector \ref{sec:model}, we consider more model building issues and discuss how the required features could be realised.

For our numerical studies, we assume that the sector that becomes strongly coupled is either supersymmetric, or such that the gaugino mass operators do not obtain large anomalous dimensions, so that the tuning with respect to the gaugino masses is reduced. To parameterise the behaviour of the visible sector in response to the hidden sector renormalisation, we take 
\begin{equation}
\tilde{\gamma}_j\left(\mu\right) = \begin{cases} 1 &\mbox{if } \Lambda_1 <\mu< \Lambda_2 \\
0 & \mbox{otherwise } \end{cases} ,
\end{equation}
where $j$ labels the chiral multiplets whose soft mass operators gain a large anomalous dimension when the hidden sector is strongly coupled, between $\Lambda_{1}$ and $\Lambda_{2}$. We have made the assumption that $\tilde{\gamma}_j >0$, in order that the RG flow decreases soft masses. This parameterisation is well motivated since at conformal fixed points of supersymmetric theories fields often have large anomalous dimensions, of order $1$ \cite{Nelson:2000sn,Intriligator:2003jj}.

The Higgs potential mass squared parameters receive a direct contribution at one loop from the charginos and the bino, and a two-loop double logarithmically enhanced contribution from the gluino through its effect on the stop. For GUT boundary conditions, the second effect is much larger. As shown in eq. \eqref{eq:1}, after running through strong coupling this contribution is strongly suppressed. The stops contribute to the Higgs mass squared at one loop through an interaction proportional to a yukawa coupling, while the other sfermions dominantly feed in through a two loop coupling. While these contributions are suppressed by the strong coupling, the sfermion soft masses are also suppressed so must be larger initially, and there is no improvement in the fine tuning with respect to these parameters.

\begin{figure}[t!]
\begin{center}
\includegraphics[width=0.635\textwidth]{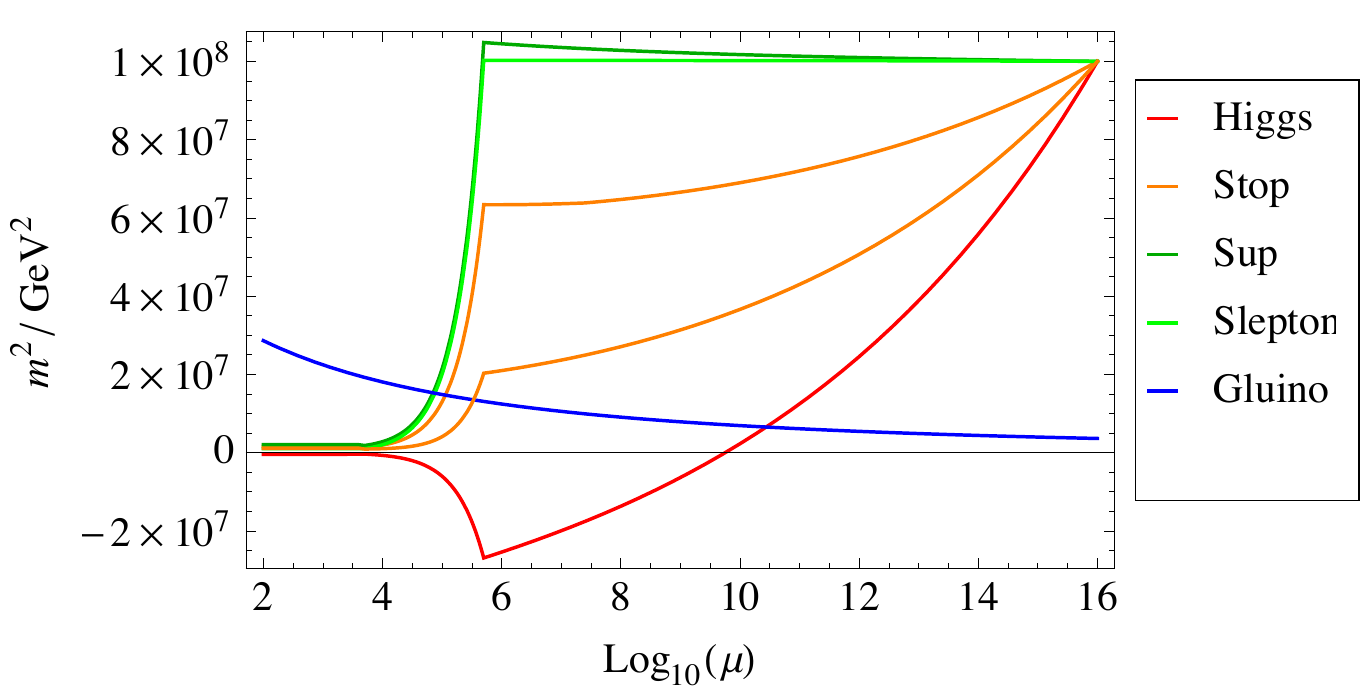} \quad \includegraphics[width=0.325 \textwidth]{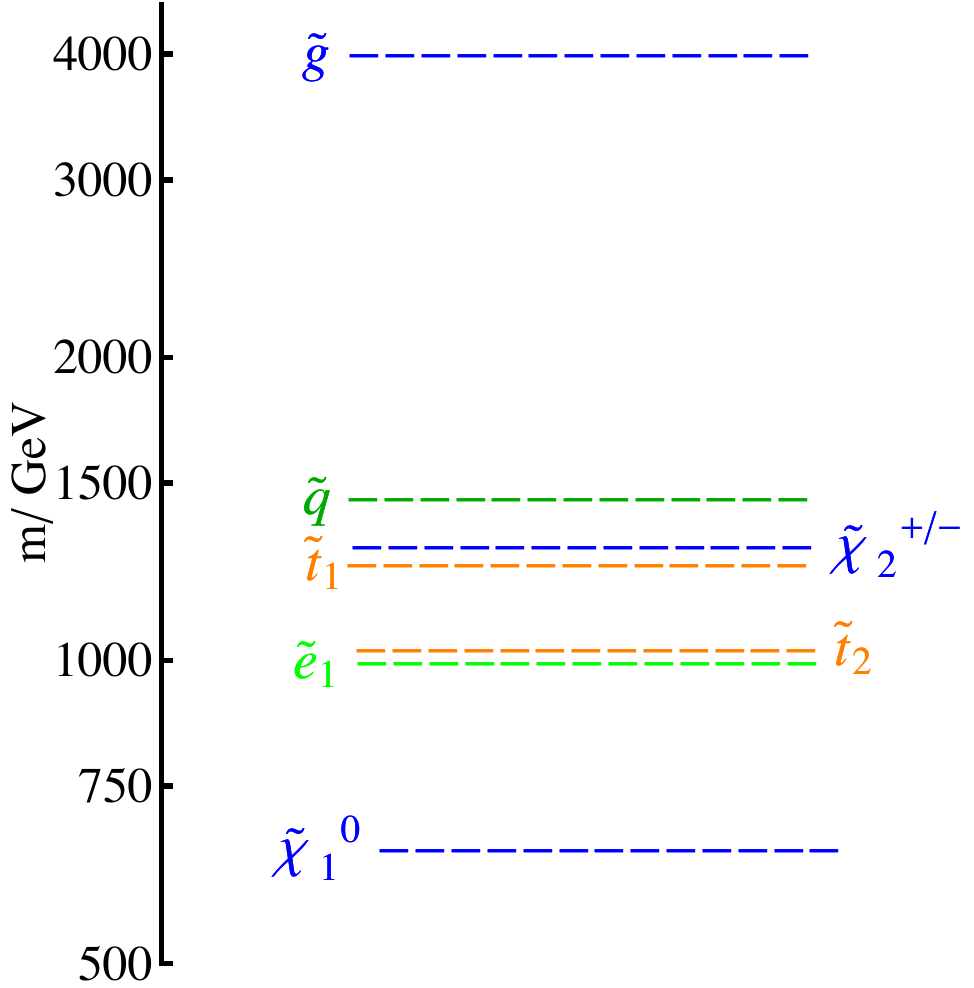}
\caption{{{\bf{Left:}}
The RG flow of the soft masses squared in the theory described in the text, assuming universal scalar soft masses at the UV boundary of the flow. The region of strong coupling is clearly visible as an approximately exponential suppression of the scalar soft masses in the region of $10^5\,\rm{GeV}$. The Higgs soft mass runs to negative coupling, driven by the terms proportional to the top Yukawa coupling in the RG equations, compatible with radiative electroweak symmetry breaking. {\bf Right:}} The weak scale soft masses in the same theory. The gluino mass is far above LHC reach without being the dominant tuning in the theory, and the sfermion masses are close to current limits.}
\label{Fig:run}
\end{center}
\end{figure} 

To study these effects more carefully, we analyse the RG flow of the MSSM in the presence of hidden sector renormalisation numerically. We include the full one-loop equations, and the dominant two-loop effects. Initially, we consider a theory with high-scale mediation. The effect of the hidden sector renormalisation is especially dramatic in this case, and the results are very similar to the low-scale mediation case, except that  models with low-scale mediation have slightly less fine tuning. Further, we assume the soft mass operators of all chiral multiplets in the theory gain a large anomalous dimension in the strong coupling region. This is completely flavour blind, and attractive in its simplicity. All the dynamics that leads to the large anomalous dimensions can be generated in the hidden sector, without any additions to the visible sector. Also, constraints on flavour changing current are automatically satisfied, even though the strong coupling region is close to the weak scale and  higher dimension operators are not strongly suppressed.

Fig. \ref{Fig:run} (left) shows the running in a typical theory, under the assumption of universal soft masses of $10^4\,\rm{GeV}$ at a mediation scale of $10^{16}\,\rm{GeV}$, and universal gaugino masses of $1400 \,\rm{GeV}$ (corresponding to a weak scale gluino mass of $4\,\rm{TeV}$), characteristic of a GUT theory. The fine tuning is not substantially altered if the initial soft masses fall into the pattern predicted by minimal gauge mediation. The period of strong coupling is taken to be between $5\times 10^5\,\rm{GeV}$ and $5\times 10^3 \,\rm{GeV}$ and is clearly visible in its effect on the scalar soft masses, while the gluino mass is unaffected. In Fig. \ref{Fig:run} (right), we show the mass spectrum obtained at the weak scale. The sfermions are close to the current experimental bound and the gauginos are far above the regions that can be efficiently probed by the LHC. 
Since the scale of mediation is high, the F-term is large, of order $\sqrt{F} \sim 10^{10}\,\rm{GeV}$.


The fine tuning with respect to the initial universal gaugino mass squared is one part in 15. This a substantial improvement over the typical tuning from a 4 TeV weak scale gluino mass with a mediation scale of $10^{16}\,\rm{GeV}$ which is in the region of one part in $600$. Taking a lower mediation scale or the region of strong coupling closer to the weak scale can lower the tuning obtained to one part in 5. The tuning from sfermions is of order one part in 70 which is comparable to that in a model without strong coupling and identical weak scale scalar masses. Assuming just an MSSM Higgs structure, $\mu$ is fixed by eq. \eqref{eq:ew} and induces a substantial tuning of roughly one part in $70$. One minor benefit for the fine tuning with respect to sfermions in models with strong coupling is that heavy gluinos slightly reduce production cross sections and consequently alleviate collider bounds on sfermions. However, even with a decoupled gluino, the limits on universal sfermion masses are in the region of $1.4\,\rm{TeV}$, which is substantial.

\begin{figure}[t!]
\begin{center}
\includegraphics[width=1\textwidth]{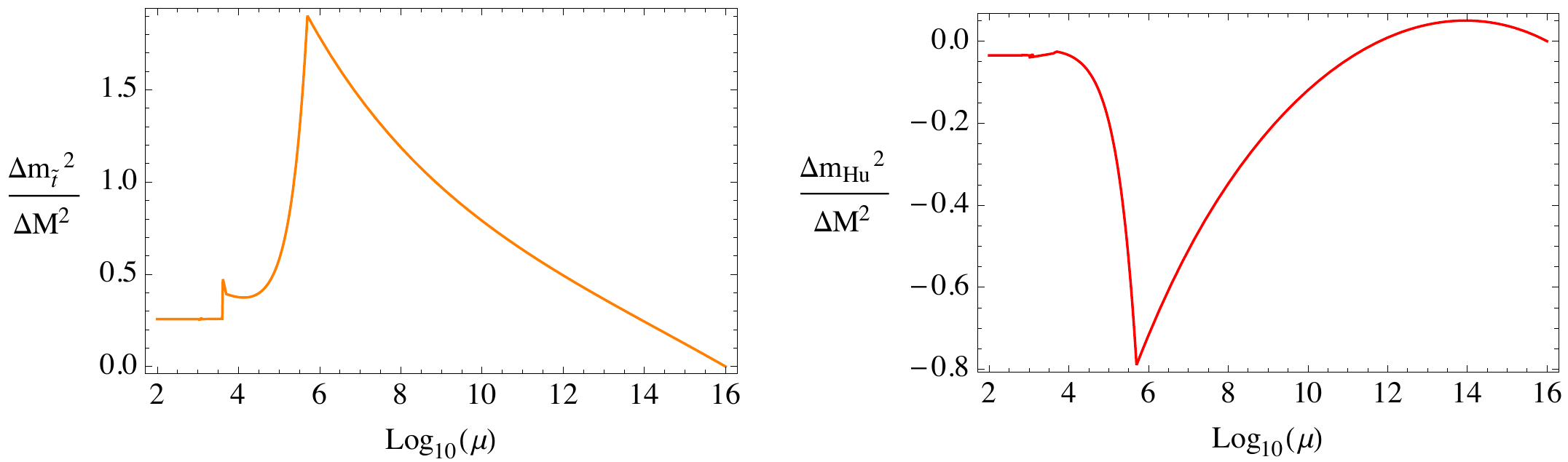}
\caption{
 \textbf{Left:} The perturbation induced in the stop soft mass, $\Delta m_{\tilde{t}}^2$, in response to a perturbation of the universal gaugino mass, $\Delta M^2$, at the UV boundary of the RG flow (taken to be $10^{16}\,\rm{GeV}$), as a function of the energy scale during the RG flow. \textbf{Right:} The perturbation induced in the up type Higgs mass, $\Delta m_{Hu}^2$, in response to the same gaugino mass perturbation. The parameters are those of the theory plotted in Fig. \ref{Fig:run} and described in the text.}
\label{Fig:resp}
\end{center}
\end{figure} 

We also plot the perturbation induced in the stop and Higgs soft masses as a result of a perturbation to the universal gaugino mass at the high scale in Fig. \ref{Fig:resp}. Initially a large perturbation in the stop soft mass is induced by the perturbation to the gluino mass (this is identical to the start of the RG flow that would be followed if it wasn't for the strong coupling). The strong coupling regime is clearly visible and heavily suppresses the perturbation to the stop mass, followed by a short period where the correction is regenerated, before the gluino is integrated out of the theory. The small near vertical drop is because the gluino is now slightly heavier and integrated out of the theory earlier resulting in less running. The Higgs mass (right panel), initially experiences a small positive perturbation as a result of the increased chargino and bino mass. After a short distance in energy scale this is overwhelmed by the two-loop contribution due to the increased mass of the stop. Again, the perturbation is strongly suppressed in the strong coupling region, and the regeneration is negligible.

There are extensions to the simplest models that reduce the tuning with respect to the sfermion masses. Of course, the pay off for this is that the couplings between the visible sector and hidden sector have to be more complicated. Consider a theory where the sfermions have flavour universal couplings to the hidden sector spurion, through ordinary gauge mediation, but the Higgs fields have additional couplings to the spurion. As a result the initial Higgs soft masses are enhanced compared to normal gauge mediation. Further, assume the extra Higgs couplings result in the Higgs soft mass operator gaining a large anomalous dimension during the strong coupling period, whereas the other operators do not. Such a structure may be obtained, for example, if the Higgs fields are charged under a new gauge symmetry which the other visible sector fields are not.\footnote{Of course, such a symmetry must be strongly broken to allow the Standard Model Yukawa couplings.} Given that a complete theory must include a mechanism to solve the $\mu / B\mu$ problems it is not unreasonable that the Higgs fields have different interactions to the other chiral multiplets.  Flavour observables are still safe since the sfermion couplings are universal. The perturbation to the Higgs mass from the sfermion masses is suppressed by the strong coupling, but the sfermion masses are not themselves suppressed, reducing the fine tuning from this sector. The initial large Higgs mass is actually useful in finding spectra where the up type Higgs soft mass is close to zero after running, reducing the required value of $\mu$ from eq. \eqref{eq:ew}, and consequently the tuning from this parameter.

\begin{figure}[t!]
\begin{center}
\includegraphics[width=0.635\textwidth]{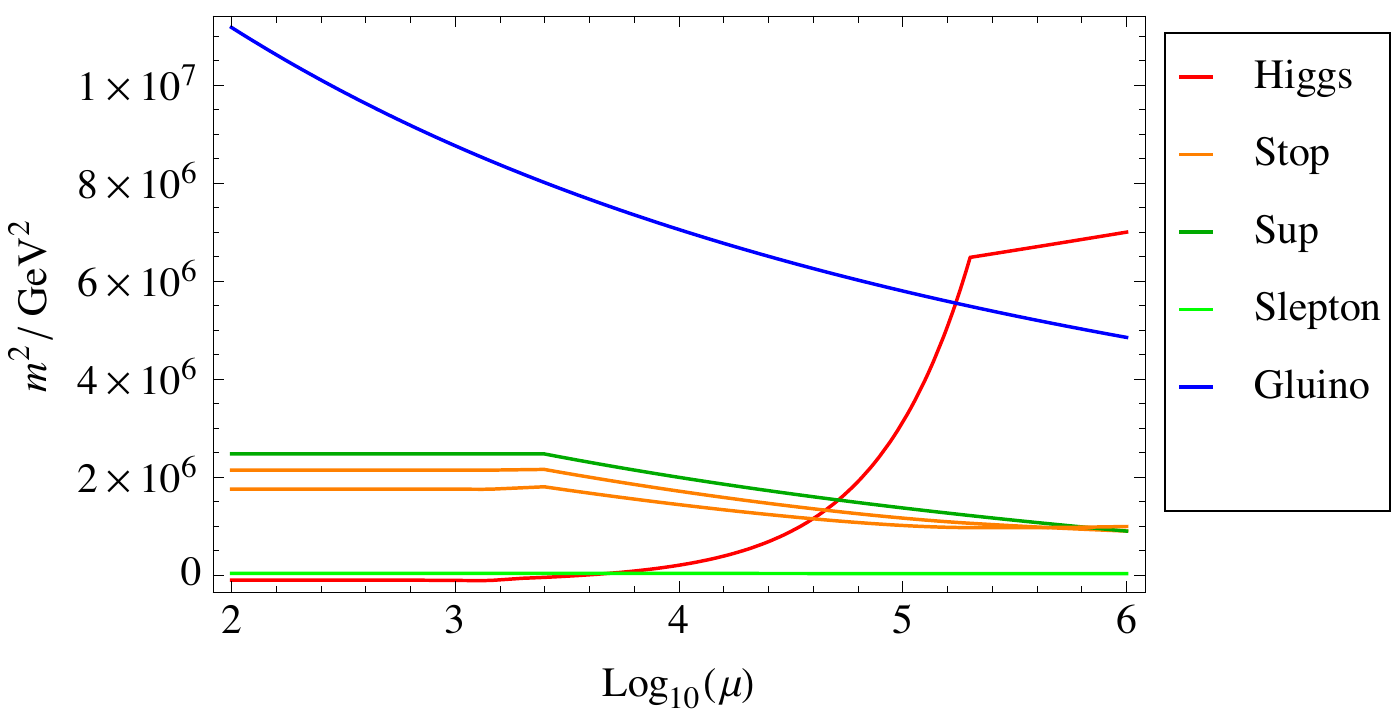} \quad \includegraphics[width=0.325\textwidth]{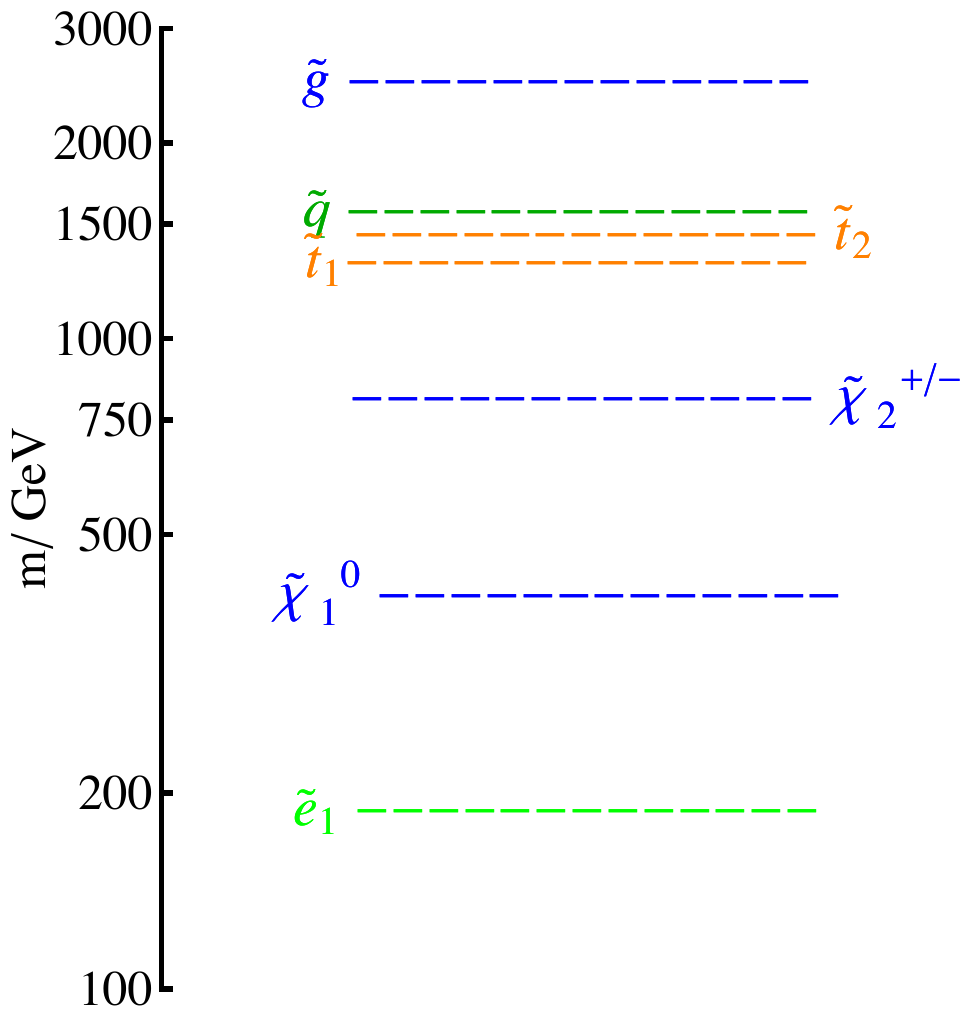}
\caption{{{\bf{Left:}}
The RG flow of the soft masses squared in the theory described in the text. Only the Higgs mass squared operator gains a large anomalous dimension in the strong coupling region, and is consequently suppressed. The mediation scale is low, and the sfermions mass ratio at the UV boundary of the RG flow is that of standard gauge mediation. The gauginos masses fall into the GUT pattern, but are assumed to be independent of the sfermion and Higgs masses. {\bf Right:}} The weak scale soft masses in the same theory. The UV boundary parameters of the theory are such that the sfermion masses are close to current LHC limits.}
\label{Fig:run2}
\end{center}
\end{figure}

In Fig. \ref{Fig:run2} (left), we plot a model with these features. The theory has a low mediation scale of $10^6 \,\rm{GeV}$, and a strong coupling region between $2\times 10^5 \,\rm{GeV}$ and $2\times 10^3 \,\rm{GeV}$. $\sqrt{F}$ is approximately $10^5 \,\rm{GeV}$, just inside the strong coupling region. The sfermion soft masses at the UV boundary of the RG flow are taken to fall into the standard gauge mediation pattern. The weak scale masses are shown in Fig. \ref{Fig:run2} (right). The gluino is at $2.5 \,\rm{TeV}$ and the squarks are in the region of $1.55 \,\rm{TeV}$, very close to current limits. For the initial parameters chosen, the Higgs mass squared just runs negative, with a weak scale value of $-\left(320\,\rm{GeV}\right)^2$, and $\mu = 313\,\rm{GeV}$. The tuning with respect to the initial gaugino mass squared is approximately one part in $15$, with respect to the initial sfermion approximately one part in $20$, and that with respect to the initial Higgs mass, and also the initial value of $\mu$ is also in the region of one part in $20$. This is a substantial improvement over a low scale gauge mediation model without hidden sector renormalisation, which typically has a tuning of one part in $\mathcal{O}\left(100\right)$ in each of these parameters.

As a entertaining alternative, it is possible that the anomalous dimension of the sfermion mass operators could actually become negative in the strong coupling regime. This would lead to an \emph{enhancement} of the sfermion masses during the strong coupling region, taking them far out of LHC reach without introducing significant fine tuning. It could be that all the sfermions receive enhanced masses, or alternatively just the first two generation sfermions might be enhanced. The later could occur, for example, if these generations are charged under an additional broken gauge group. It is possible such a structure could be linked to the fermion mass heirachy, in the style of the classic `natural' SUSY spectra \cite{Dimopoulos:1995mi,Cohen:1996vb}.\footnote{A similar mechanism for generating a `natural' spectrum has been studied in \cite{Cohen:2012rm}.} This breaking of a gauged flavour group, could even trigger the supersymmetry breaking sector to run into strong coupling, especially since the SU(2) structures which arise in the model of \cite{Craig:2012di} often appear in interesting candidates for conformal theories.\footnote{We are grateful to Matthew McCullough for this observation.}

Assuming all sfermion soft masses are enhanced, it is possible to obtain a model with a weak scale gluino mass of $3\,\rm{TeV}$, and sfermion masses in the region of $7\,\rm{TeV}$, with an associated tunings of only one part in 25 and one part in 50 with respect to the sfermion and gluino masses respectively.  Overall, the tuning of the theory is comparable to the most natural traditional models compatible with collider bounds, even though the scalars are very heavy with a spectrum reminiscent of mini-split supersymmetry \cite{Arvanitaki:2012ps}. Since the stop masses are much increased relative to traditional models, it is far easier to obtain a physical Higgs mass of $125\,\rm{GeV}$ compared to conventional models. The one loop corrections to the physical Higgs mass are given by (assuming vanishing A-terms)
\begin{equation}
\Delta \left( m_{h0}^2 \right) = \frac{3 y_t^2}{4 \pi^2}  m_t^2 \log \left(\frac{m_{\tilde{t}}^2}{m_t^2} \right) .
\end{equation}
Notably, this correction is cut off by the mass of the stops, not the UV boundary of the RG flow, since it is the quartic Higgs coupling which is important. Therefore, the strong coupling region (which is typically above the scale of the stops) does not alter the form of the correction to the physical Higgs mass.


While we have studied hidden sectors with rather dramatic effects on the RG flow, it is also plausible that the fine tuning may be reduced substantially even in a model where the hidden sector does not become strongly coupled. For example, a weakly coupled hidden sector could modify the visible sector running in such a way that additional cancellations between the various contributions to the Higgs soft mass appear. This is somewhat analogously to `focus point' models \cite{Kaminska:2013mya}, and relies on the careful analysis of different tuning measures emphasised in \cite{Baer:2013gva}. It would be interesting to find examples of theories where this could occur.


\section{Model building}\label{sec:model}
We now return to discuss models which can lead to the features assumed in the previous section. Finding and studying explicit examples of non-supersymmetric theories which pass close to an interacting conformal fixed point is hard, however several examples are believed to exist. In fact, these types of model have been studied extensively in the context of walking technicolor  \cite{Sannino:2004qp,Dietrich:2005jn}. It is somewhat easier to find supersymmetric theories with the appropriate dynamics, see for example \cite{Poland:2011kg}.


\subsubsection*{Low scale mediation}
The simplest implementation of hidden sector renormalisation reducing fine tuning arises in models with low scale mediation. In such models, SUSY breaking occurs at approximately the same scale as the strong coupling region, and there is the potential to link these two events, for example running to strong coupling could trigger spontaneous supersymmetry breaking as happens in a number of known models.

Typically, due to loop factors that arise in the gauge mediation to the visible sector, $\sqrt{F}$ must be above the weak scale and the lower limit of the strong coupling region. The hidden sector, and SUSY breaking multiplet, can easily remain dynamical below this scale if some fields in the hidden sector have masses suppressed by loop factors or small coupling constants. For example, the dominant F-term in the theory can arise as an expectation value of a scalar field which receives a mass only at loop order. Additionally SUSY breaking masses that appear in the hidden sector can be somewhat removed from $\sqrt{F}$. If this is the case, the scale of these masses may be close to the end of the strong coupling region. Consequently, at least some of the interactions in the strong coupling region can remain approximately supersymmetric, and results from holomorphy may remain accurate. Alternatively, there may exist mediation mechanisms which do not lead to loop factors, so that $\sqrt{F}$ can be close to the weak scale, and the hidden sector remains supersymmetric during strong coupling.  Conversely, the hidden sector may be non-supersymmetric for some or all of the strong coupling region, which is not problematic but does require the extra assumption that the anomalous dimensions of the  gaugino mass operators are small. 

In models with low scale mediation, the RG flow of the SM gauge couplings is not necessarily altered; the strong coupling region may only affect matter in the SUSY breaking sector, which is uncharged under the SM gauge groups. SUSY breaking messengers, charged under the SM gauge group, might not experience strong coupling depending on the dynamics of their couplings to the hidden sector. Even if the RG flows are modified, gauge unification can be maintained provided the matter content and couplings of the messenger sector are GUT compatible.

To clarify some of these issues we consider a well known example of a dynamical SUSY breaking sector, the ISS model \cite{Intriligator:2006dd}. Although we do not provide a full model (for example, a mediation mechanism) or calculation of anomalous dimensions, this is a good candidate for a theory which might remain strongly coupled for an extended energy range if the theory is close to the edge of the `conformal window', that is, $N_f$ is close to $\frac{3}{2}N_c$ \cite{Poland:2011kg}. Here we give only a brief overview of the ISS model, more details may be found in \cite{Intriligator:2006dd}. The UV (asymptotically free) description consists of SQCD with massive quarks, of mass $m$, in the window $N_c + 1 \leq N_f < \frac{3}{2} N_c$, where $N_f$ is the number of flavours and $N_c$ is the number of colours. The theory runs to strong coupling at a scale $\Lambda$, which by assumption $\gg m$. Below the strong coupling scale, the theory is described by a dual magnetic theory. This has gauge group $SU\left(N_f-N_c\right)$, field content consisting of dual quarks $q,\,\tilde{q}$ and the electric theory's meson $\Phi_i^j$, and a superpotential
\begin{equation}
W = h \rm{Tr}\left( q \Phi \tilde{q}\right) - \emph{h} \mu^2 \rm{Tr}\left(\Phi\right) .
\end{equation}
Here, $h$ is a coupling constant in the theory and $\mu$ is a mass scale set by $h^{1/2} \mu \sim \left(\Lambda m\right)^{1/2} \ll \Lambda$. The magnetic theory breaks SUSY since a rank condition means that not all F-terms can be set to zero. The supersymmetry breaking that results is given by
\begin{equation}
\sqrt{\sum F} \sim \left(N_f - N_c\right)^{1/2} h^{1/2} \mu \ll \Lambda. 
\end{equation}
Typical supersymmetric and non-supersymmetric masses in the theory are $\sim h \mu$, while directions which are pseudomoduli (including the scalar components of the multiplets that obtain F-terms) only obtain masses at one loop of size $\sim h^2 \mu$ which are much less than the soft masses if $h$ is small.

\begin{figure}
\centering
\begin{minipage}{.47\textwidth}
  \centering
  \includegraphics[width=0.85\textwidth]{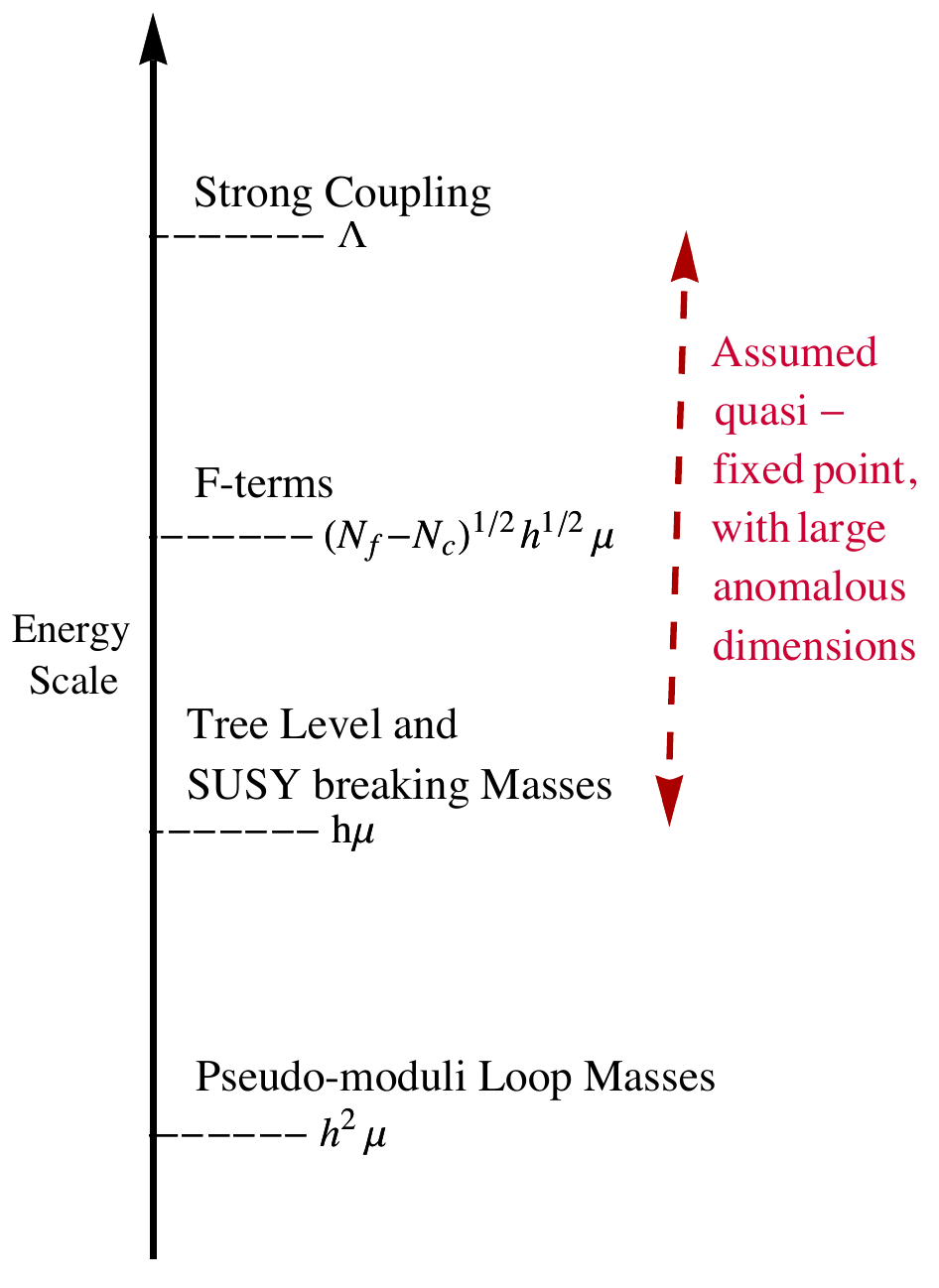}
  \captionof{figure}{The mass scales in the ISS model, which may (once a complete model, including messenger sector, is specified) be a candidate for a theory with low scale mediation and significant hidden sector renormalisation. The parameters are as defined in the text.}
  \label{Fig:issscales}
\end{minipage}%
\qquad
\begin{minipage}{.47\textwidth}
  \centering
  \includegraphics[width=0.49\textwidth]{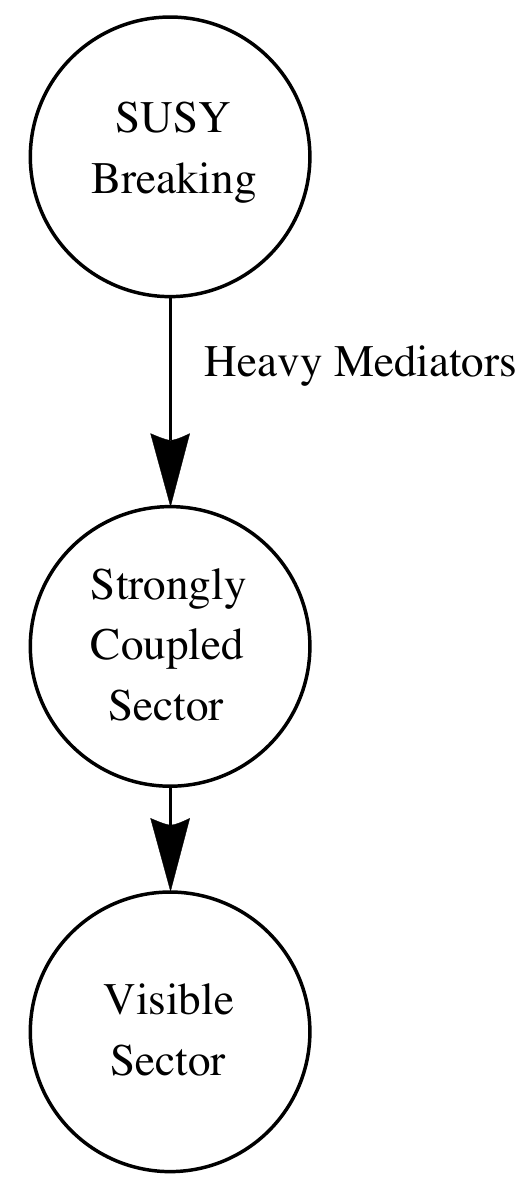}
  \captionof{figure}{ A schematic setup which could lead to hidden sector renormalisation, near the weak scale, in a model with high scale mediation. The strong coupling region may be approximately supersymmetric depending on the interactions between the sectors.}
  \label{Fig:highscale}
\end{minipage}
\end{figure}


Not all constants are set by holomorphy, as an example we suppose that the theory happens to have small  $h$. In this case the heirachy of masses is as shown in Fig. \ref{Fig:issscales}. The theory has a separation between the onset of the strong coupling region, the scale of the F-terms in the theory and the masses of the fields. As is seen there, supersymmetry breaking occurs somewhat after the beginning of the strong coupling region. The masses of states in the sector are suppressed relative to the SUSY breaking scale, and therefore the theory remains dynamical until these masses are reached. Additionally, above the scale of the soft masses the theory some interactions of the theory may remain approximately supersymmetric, even though the energy scale is below $\sqrt{F}$. Below the soft masses the theory is non-supersymmetric but remains dynamical until the masses of the lightest states are reached.

\subsubsection*{High scale mediation}
It is also possible to build models with high scale mediation. This case is slightly different, since the SUSY breaking sector, which is typically dynamical only for a few orders of magnitude below $\sqrt{F}$, cannot be the strong coupling sector. However, hidden sector renormalisation can reduce fine tuning if the messenger sector of the theory is more complicated than usual, and becomes strongly coupled near the weak scale. While the setups involved may seem more contrived than the low scale case, they are still interesting to consider.

Suppose the theory is as shown in Fig. \ref{Fig:highscale}. The visible sector is coupled to the supersymmetry breaking sector indirectly, through a sector with light states which themselves couple to the SUSY breaking sector through heavy mediators. If the sector containing light states becomes strongly coupled, the visible sector soft mass operators still gain large anomalous dimensions. The strong coupling sector is supersymmetric until a scale $\frac{F}{M_{med}}$, which can be close to the visible sector soft masses and the weak scale. Depending on the model, the strong coupling region may or may not be supersymmetric. A particularly attractive scenario is if the soft masses in the light messenger sector cause the sector to leave the strong coupling regime, so that both are close to the weak scale in a correlated fashion.

\begin{figure}[t!]
\begin{center}
\includegraphics[width=0.6\textwidth]{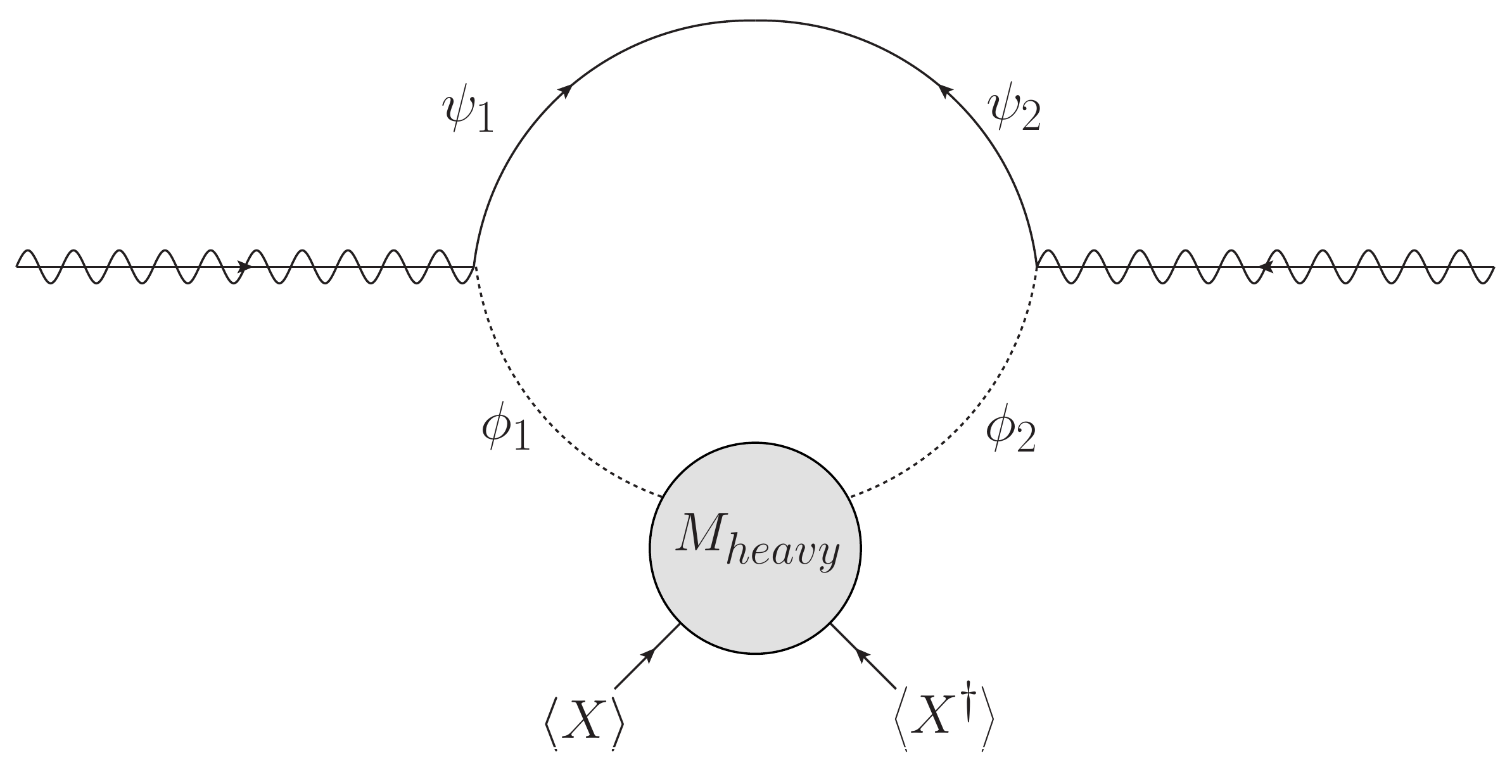}
\caption{
A diagram that can lead to visible sector gaugino masses in a model with high scale mediation. The fields $\psi_{1,2}$ and $\phi_{1,2}$ are the fermions and scalars in the messenger chiral multiplets $\Psi_{1,2}$ and have masses near the TeV scale. Even though the mediation scale is high, the messenger multiplets may remain approximately supersymmetric and dynamical until close to the weak scale. Consequently hidden sector renormalisation can reduce the theory's EW fine tuning. Visible sector scalar soft masses squared are be generated through higher loop order diagrams.}
\label{Fig:high}
\end{center}
\end{figure} 

For example, suppose a theory contains messenger chiral superfields, $\Psi_1$ and $\Psi_2$, charged under the SM gauge groups, with a supersymmetric Dirac mass, $m_{light}$, of order a few TeV,
\begin{equation}
\mathcal{L}\supset \int d^2\theta m_{light} \Psi_1 \Psi_2 .
\end{equation}
The SUSY breaking sector is parameterised by a spurion $X$, which obtains an F-term $\left<X\right> = \theta^2 F$. Further, assume the theory contains heavy fields with typical masses of $M_{heavy}$, which couple the messengers to the SUSY breaking sector for example through some new gauge group, although we do not specify the interactions or field content. SUSY breaking masses of typical size $\frac{F}{M_{heavy}}$ are induced in the messenger scalars. Soft masses of comparable size are generated in the visible sector, through diagrams of the form of Fig. \ref{Fig:high} for gauginos and similar diagrams for the scalar masses. Depending on the details of the interactions in the hidden sector the scale of mediation, where visible sector soft masses are generated, is still typically in the region of $M_{heavy}$.\footnote{This is  similar to ordinary gauge mediation, where visible sector scalar soft masses are generated at the messenger mass scale, despite coupling to the messengers through gauginos and gauge bosons which are light.} However, since the messenger multiplets have masses near the weak scale, they remain dynamical down to low energy scales. If their interactions are such that they run into a strong coupling regime near the weak scale (for example, if they are charged under some additional gauge group that becomes strongly coupled), the visible sector soft mass operators can obtain a large anomalous dimensions, leading to hidden sector renormalisation.

Of course, it would be highly desirable to have an example of a complete theory with the required properties, and it may be possible to study suitable theories using the techniques of general gauge mediation \cite{Meade:2008wd}, however here we simply assume that such a model can be realised. In these models the RG flow of the SM gauge couplings will be altered, since the messengers, which are charged under the SM gauge groups, gain large anomalous dimensions which affect the beta functions through eq. \eqref{eq:nsvz}.\footnote{This will also slightly modify the RG flow of the physical gaugino masses, but does not have a significant effect on the fine tuning of the theory.} However gauge unification can persist if the couplings and matter content respect an underlying GUT structure. Models with high scale mediation have an additional benefit that the lightest supersymmetric particle can be a rather good dark matter candidate. This is in contrast to the very light gravitino that accompanies models of very low scale mediation, which (although a possibility) is often a rather poor dark matter candidate. Consequently, it is interesting that the models we study allow for high scale mediation, without the enormous fine tuning that usually accompanies this.


\subsection*{Other model building possibilities}
While we have focused on the benefit of hidden sector renormalisation for fine tuning, there are also potential benefits for building models of SUSY breaking. A very common issue encountered in theories with dynamical SUSY breaking is generating weak scale gaugino masses which are not suppressed relative to sfermion masses, leading to unacceptably large tuning if collider constraints are to be satisfied. Suppressed gaugino masses appear because an R-symmetry is a necessary condition for SUSY to be spontaneously broken \cite{Nelson:1993nf}. Such an R-symmetry however forbids gaugino masses, and even if it is spontaneously broken it is often not broken strongly enough. Alternatively, a metastable SUSY breaking vacua may be obtained in a sector with an approximate R-symmetry. Even here, obtaining both a sufficiently long lived vacua and heavy enough gauginos, makes model building challenging \cite{Abel:2009ze}. The theories initially studied, with all chiral multiplets getting a large anomalous dimension in the strong coupling region allows for scalars to start off heavy, yet end up lighter than the gauginos at the weak scale, alleviating this problem without introducing fine tuning.


Finally, an attractive alternative to the Majorana gauginos considered so far are Dirac gauginos. A complete discussion can be found in  \cite{Fox:2002bu,Kikuchi:2008ws,Carpenter:2010as,Benakli:2011vb,Kribs:2012gx} , but briefly, provided the mediation takes place through a so-called supersoft operator, the gauginos do not contribute to the sfermion or Higgs masses until the energy scale is below the mass of the sgauge states (these are the new scalar fields in the adjoint of the gauge group which complete the $\mathcal{N}=2$ gauge multiplet). Typically, the sgauge particles are close in mass to the gauginos, and so there is very little time for the gauginos to alter the running of the Higgs and tuning can be dramatically reduced. If Dirac gauginos are combined with hidden sector renormalisation,  in such a way that the theory is in strong coupling for most of the energy region between the masses of the sgauge and gauginos, there is a double suppression of the tuning. It is possible to obtain gluino masses of order $10\,\rm{TeV}$ without any appreciable fine tuning. While it is remarkable that there may be such little tuning from this sector of the theory, this does not greatly improve the overall fine tuning of the theory. Even with the extended set-ups to reduce sfermion fine tuning discussed in this section,  the overall tuning of the theory is still typically one part in 20 due to the initial Higgs soft mass and $\mu$ parameter required to obtain the correct EW VEV.

There is an interesting additional feature in models with Dirac gauginos that may be helpful with model building, the visible sector retains an R-symmetry. Suppose the theory has an unusual mediation structure such that $\sqrt{F}$ is below the strong coupling scale. Then the hidden sector passes close to a superconformal fixed point, and necessarily has an exact R-symmetry \cite{Nelson:2000sn,Nelson:2001mq}. This can be identified with the R-symmetry in the visible sector, allowing the anomalous dimensions of operators to be evaluated exactly.


\section{Concluding remarks}\label{sec:con}
To conclude, we have shown that the fine tuning of supersymmetric theories compatible with LHC bounds can be significantly reduced by hidden sector renormalisation in models where the SUSY breaking sector runs through an extended period of strong coupling not far from the weak scale. In the simplest implementation, the fine tuning with respect to the UV boundary gaugino mass may be reduced from one part in $\mathcal{O}\left(1000\right)$ to one part in $10$, however there is no improvement in the tuning with respect to the sfermion masses. More complicated models, where the Higgs has additional couplings to the SUSY breaking spurion, may reduce the tuning with respect to all parameters of the theory to the region of one part in $20$. In this work we have simply parameterised the hidden sector through assumed values of anomalous dimensions. It is clearly an important challenge to build realistic models of the hidden sector with the appropriate dynamics.

\section*{Acknowledgements}
I am grateful to  Ulrich Haisch, John March-Russell, Matthew McCullough, Andrew Powell, and Stephen West for very useful discussions  and comments on a draft of the manuscript, and an anonymous JHEP referee for pointing out the importance of \cite{Cohen:2006qc} for the calculation of fine tuning during the review process of \cite{Hardy:2013ywa}. 

\bibliographystyle{JHEP}
\bibliography{HRGrefs}

\providecommand{\href}[2]{#2}\begingroup\raggedright\begin{thebibliography}{10}

\bibitem{Nelson:2000sn}
A.~E. Nelson and M.~J. Strassler, {\it Suppressing flavor anarchy},  {\em JHEP}
  {\bf 09} (2000) 030, [\href{http://xxx.lanl.gov/abs/hep-ph/0006251}{{\tt
  hep-ph/0006251}}].

\bibitem{Dine:2004dv}
M.~Dine, P.~Fox, E.~Gorbatov, Y.~Shadmi, Y.~Shirman, et~al., {\it {Visible
  effects of the hidden sector}},  {\em Phys.Rev.} {\bf D70} (2004) 045023,
  [\href{http://xxx.lanl.gov/abs/hep-ph/0405159}{{\tt hep-ph/0405159}}].

\bibitem{Cohen:2006qc}
A.~G. Cohen, T.~S. Roy, and M.~Schmaltz, {\it Hidden sector renormalization of
  mssm scalar masses},  {\em JHEP} {\bf 02} (2007) 027,
  [\href{http://xxx.lanl.gov/abs/hep-ph/0612100}{{\tt hep-ph/0612100}}].

\bibitem{Murayama:2007ge}
H.~Murayama, Y.~Nomura, and D.~Poland, {\it {More visible effects of the hidden
  sector}},  {\em Phys.Rev.} {\bf D77} (2008) 015005,
  [\href{http://xxx.lanl.gov/abs/0709.0775}{{\tt arXiv:0709.0775}}].

\bibitem{Abe:2007ki}
H.~Abe, T.~Kobayashi, and Y.~Omura, {\it {Metastable supersymmetry breaking
  vacua from conformal dynamics}},  {\em Phys.Rev.} {\bf D77} (2008) 065001,
  [\href{http://xxx.lanl.gov/abs/0712.2519}{{\tt arXiv:0712.2519}}].

\bibitem{Cho:2008fr}
H.~Y. Cho, {\it {Constraints of the B(mu) / mu solution due to the hidden
  sector renormalization}},  {\em JHEP} {\bf 0807} (2008) 069,
  [\href{http://xxx.lanl.gov/abs/0802.1145}{{\tt arXiv:0802.1145}}].

\bibitem{Kawamura:2008bf}
Y.~Kawamura, T.~Kinami, and T.~Miura, {\it {Superparticle Sum Rules in the
  presence of Hidden Sector Dynamics}},  {\em JHEP} {\bf 0901} (2009) 064,
  [\href{http://xxx.lanl.gov/abs/0810.3965}{{\tt arXiv:0810.3965}}].

\bibitem{Campbell:2008tt}
B.~A. Campbell, J.~Ellis, and D.~W. Maybury, {\it {Observing The Hidden
  Sector}},  \href{http://xxx.lanl.gov/abs/0810.4877}{{\tt arXiv:0810.4877}}.

\bibitem{Perez:2008ng}
G.~Perez, T.~S. Roy, and M.~Schmaltz, {\it {Phenomenology of SUSY with scalar
  sequestering}},  {\em Phys.Rev.} {\bf D79} (2009) 095016,
  [\href{http://xxx.lanl.gov/abs/0811.3206}{{\tt arXiv:0811.3206}}].

\bibitem{Craig:2009rk}
N.~J. Craig and D.~Green, {\it {On the Phenomenology of Strongly Coupled Hidden
  Sectors}},  {\em JHEP} {\bf 0909} (2009) 113,
  [\href{http://xxx.lanl.gov/abs/0905.4088}{{\tt arXiv:0905.4088}}].

\bibitem{Arai:2010ds}
M.~Arai, S.~Kawai, and N.~Okada, {\it {A Gauge mediation scenario with hidden
  sector renormalization in MSSM}},  {\em Phys.Rev.} {\bf D81} (2010) 035022,
  [\href{http://xxx.lanl.gov/abs/1001.1509}{{\tt arXiv:1001.1509}}].

\bibitem{Arai:2010qe}
M.~Arai, S.~Kawai, and N.~Okada, {\it {Renormalization effects on the MSSM from
  a calculable model of a strongly coupled hidden sector}},  {\em Phys.Rev.}
  {\bf D84} (2011) 075002, [\href{http://xxx.lanl.gov/abs/1011.3998}{{\tt
  arXiv:1011.3998}}].

\bibitem{Terao:2001jw}
H.~Terao, {\it {Renormalization group for soft SUSY breaking parameters and
  MSSM coupled with superconformal field theories}},
  \href{http://xxx.lanl.gov/abs/hep-ph/0112021}{{\tt hep-ph/0112021}}.

\bibitem{Kobayashi:2004pu}
T.~Kobayashi and H.~Terao, {\it {Suppressed supersymmetry breaking terms in the
  Higgs sector}},  {\em JHEP} {\bf 0407} (2004) 026,
  [\href{http://xxx.lanl.gov/abs/hep-ph/0403298}{{\tt hep-ph/0403298}}].

\bibitem{Terao:2007pm}
H.~Terao, {\it {Higgs and top quark coupled with a conformal gauge sector}},
  \href{http://xxx.lanl.gov/abs/0705.0443}{{\tt arXiv:0705.0443}}.

\bibitem{Cohen:2012rm}
T.~Cohen, A.~Hook, and G.~Torroba, {\it {An Attractor for Natural
  Supersymmetry}},  {\em Phys.Rev.} {\bf D86} (2012) 115005,
  [\href{http://xxx.lanl.gov/abs/1204.1337}{{\tt arXiv:1204.1337}}].

\bibitem{Hardy:2013ywa}
E.~Hardy, {\it {Is Natural SUSY Natural?}},
  \href{http://xxx.lanl.gov/abs/1306.1534}{{\tt arXiv:1306.1534}}.

\bibitem{Arvanitaki:2013yja}
A.~Arvanitaki, M.~Baryakhtar, X.~Huang, K.~Van~Tilburg, and G.~Villadoro, {\it
  {The Last Vestiges of Naturalness}},
  \href{http://xxx.lanl.gov/abs/1309.3568}{{\tt arXiv:1309.3568}}.

\bibitem{Fox:2002bu}
P.~J. Fox, A.~E. Nelson, and N.~Weiner, {\it {Dirac gaugino masses and
  supersoft supersymmetry breaking}},  {\em JHEP} {\bf 0208} (2002) 035,
  [\href{http://xxx.lanl.gov/abs/hep-ph/0206096}{{\tt hep-ph/0206096}}].

\bibitem{Higashijima:2003et}
K.~Higashijima and E.~Itou, {\it {Unitarity bound of the wave function
  renormalization constant}},  {\em Prog.Theor.Phys.} {\bf 110} (2003)
  107--114, [\href{http://xxx.lanl.gov/abs/hep-th/0304047}{{\tt
  hep-th/0304047}}].

\bibitem{Baer:2013bba}
H.~Baer, V.~Barger, and M.~Padeffke-Kirkland, {\it {Electroweak versus high
  scale finetuning in the 19-parameter SUGRA model}},  {\em Phys.Rev.} {\bf
  D88} (2013) 055026, [\href{http://xxx.lanl.gov/abs/1304.6732}{{\tt
  arXiv:1304.6732}}].

\bibitem{Baer:2013gva}
H.~Baer, V.~Barger, and D.~Mickelson, {\it {How conventional measures
  overestimate electroweak fine-tuning in supersymmetric theory}},
  \href{http://xxx.lanl.gov/abs/1309.2984}{{\tt arXiv:1309.2984}}.

\bibitem{Intriligator:2003jj}
K.~A. Intriligator and B.~Wecht, {\it {The Exact superconformal R symmetry
  maximizes a}},  {\em Nucl.Phys.} {\bf B667} (2003) 183--200,
  [\href{http://xxx.lanl.gov/abs/hep-th/0304128}{{\tt hep-th/0304128}}].

\bibitem{Dimopoulos:1995mi}
S.~Dimopoulos and G.~Giudice, {\it {Naturalness constraints in supersymmetric
  theories with nonuniversal soft terms}},  {\em Phys.Lett.} {\bf B357} (1995)
  573--578, [\href{http://xxx.lanl.gov/abs/hep-ph/9507282}{{\tt
  hep-ph/9507282}}].

\bibitem{Cohen:1996vb}
A.~G. Cohen, D.~Kaplan, and A.~Nelson, {\it {The More minimal supersymmetric
  standard model}},  {\em Phys.Lett.} {\bf B388} (1996) 588--598,
  [\href{http://xxx.lanl.gov/abs/hep-ph/9607394}{{\tt hep-ph/9607394}}].

\bibitem{Craig:2012di}
N.~Craig, M.~McCullough, and J.~Thaler, {\it {Flavor Mediation Delivers Natural
  SUSY}},  {\em JHEP} {\bf 1206} (2012) 046,
  [\href{http://xxx.lanl.gov/abs/1203.1622}{{\tt arXiv:1203.1622}}].

\bibitem{Arvanitaki:2012ps}
A.~Arvanitaki, N.~Craig, S.~Dimopoulos, and G.~Villadoro, {\it {Mini-Split}},
  {\em JHEP} {\bf 1302} (2013) 126,
  [\href{http://xxx.lanl.gov/abs/1210.0555}{{\tt arXiv:1210.0555}}].

\bibitem{Kaminska:2013mya}
A.~Kaminska, G.~G. Ross, and K.~Schmidt-Hoberg, {\it {Non-universal gaugino
  masses and fine tuning implications for SUSY searches in the MSSM and the
  GNMSSM}},  \href{http://xxx.lanl.gov/abs/1308.4168}{{\tt arXiv:1308.4168}}.

\bibitem{Sannino:2004qp}
F.~Sannino and K.~Tuominen, {\it {Orientifold theory dynamics and symmetry
  breaking}},  {\em Phys.Rev.} {\bf D71} (2005) 051901,
  [\href{http://xxx.lanl.gov/abs/hep-ph/0405209}{{\tt hep-ph/0405209}}].

\bibitem{Dietrich:2005jn}
D.~D. Dietrich, F.~Sannino, and K.~Tuominen, {\it {Light composite Higgs from
  higher representations versus electroweak precision measurements: Predictions
  for CERN LHC}},  {\em Phys.Rev.} {\bf D72} (2005) 055001,
  [\href{http://xxx.lanl.gov/abs/hep-ph/0505059}{{\tt hep-ph/0505059}}].

\bibitem{Poland:2011kg}
D.~Poland and D.~Simmons-Duffin, {\it {N=1 SQCD and the Transverse Field Ising
  Model}},  {\em JHEP} {\bf 1202} (2012) 009,
  [\href{http://xxx.lanl.gov/abs/1104.1425}{{\tt arXiv:1104.1425}}].

\bibitem{Intriligator:2006dd}
K.~A. Intriligator, N.~Seiberg, and D.~Shih, {\it {Dynamical SUSY breaking in
  meta-stable vacua}},  {\em JHEP} {\bf 0604} (2006) 021,
  [\href{http://xxx.lanl.gov/abs/hep-th/0602239}{{\tt hep-th/0602239}}].

\bibitem{Meade:2008wd}
P.~Meade, N.~Seiberg, and D.~Shih, {\it {General Gauge Mediation}},  {\em
  Prog.Theor.Phys.Suppl.} {\bf 177} (2009) 143--158,
  [\href{http://xxx.lanl.gov/abs/0801.3278}{{\tt arXiv:0801.3278}}].

\bibitem{Nelson:1993nf}
A.~E. Nelson and N.~Seiberg, {\it {R symmetry breaking versus supersymmetry
  breaking}},  {\em Nucl.Phys.} {\bf B416} (1994) 46--62,
  [\href{http://xxx.lanl.gov/abs/hep-ph/9309299}{{\tt hep-ph/9309299}}].

\bibitem{Abel:2009ze}
S.~A. Abel, J.~Jaeckel, and V.~V. Khoze, {\it {Gaugino versus Sfermion Masses
  in Gauge Mediation}},  {\em Phys.Lett.} {\bf B682} (2010) 441--445,
  [\href{http://xxx.lanl.gov/abs/0907.0658}{{\tt arXiv:0907.0658}}].

\bibitem{Kikuchi:2008ws}
T.~Kikuchi, {\it {A Solution to the little hierarchy problem in a partly N=2
  extension of the MSSM}},  \href{http://xxx.lanl.gov/abs/0812.2569}{{\tt
  arXiv:0812.2569}}.

\bibitem{Carpenter:2010as}
L.~M. Carpenter, {\it {Dirac Gauginos, Negative Supertraces and Gauge
  Mediation}},  {\em JHEP} {\bf 1209} (2012) 102,
  [\href{http://xxx.lanl.gov/abs/1007.0017}{{\tt arXiv:1007.0017}}].

\bibitem{Benakli:2011vb}
K.~Benakli, {\it {Dirac Gauginos: A User Manual}},  {\em Fortsch.Phys.} {\bf
  59} (2011) 1079--1082, [\href{http://xxx.lanl.gov/abs/1106.1649}{{\tt
  arXiv:1106.1649}}].

\bibitem{Kribs:2012gx}
G.~D. Kribs and A.~Martin, {\it {Supersoft Supersymmetry is Super-Safe}},  {\em
  Phys.Rev.} {\bf D85} (2012) 115014,
  [\href{http://xxx.lanl.gov/abs/1203.4821}{{\tt arXiv:1203.4821}}].

\bibitem{Nelson:2001mq}
A.~E. Nelson and M.~J. Strassler, {\it Exact results for supersymmetric
  renormalization and the supersymmetric flavor problem},  {\em JHEP} {\bf 07}
  (2002) 021, [\href{http://xxx.lanl.gov/abs/hep-ph/0104051}{{\tt
  hep-ph/0104051}}].

\end{thebibliography}\endgroup

\end{document}